\documentstyle[12pt]{article}

\input epsf

\newcommand{\resection}[1]{\setcounter{equation}{0}\section{#1}}

\newcommand{\beqn}{\begin{equation}}
\newcommand{\eeqn}{\end{equation}}
\newcommand{\beqa}{\begin{eqnarray}}
\newcommand{\eeqa}{\end{eqnarray}}
\newcommand{\nn}{\nonumber \\}
\newcommand{\sh}{{\sinh}}
\newcommand{\ch}{{\cosh}}

\begin{document}

\begin{titlepage}

\begin{flushleft}
\setlength{\baselineskip}{13pt}
Yukawa Institute Kyoto \hfill YITP-97-33 \\
		       \hfill June 1997
\end{flushleft}

\vspace*{\fill}
\begin{center}
{\LARGE{\bf  $q$-deformed Coxeter element 
            in Non-simply-laced Affine Toda Field Theories}} \\
\vfill
{\sc Takeshi Oota}
\footnote{e-mail: {\tt toota@yukawa.kyoto-u.ac.jp}}\\[2em]

{\sl Yukawa Institute for Theoretical Physics, Kyoto
           University, Kyoto 606-01, Japan}\\

\vfill
ABSTRACT
\end{center}
\begin{quote}
The Lie algebraic structures of the $S$-matrices for the affine Toda
field theories based on the dual pairs $(X_N^{(1)}, Y_M^{(l)})$ are
discussed. For the non-simply-laced horizontal subalgebra $X_N$ and
the simply-laced horizontal subalgebra $Y_M$, we introduce
a ``$q$-deformation'' of a Coxeter element and 
a ``$p$-deformation'' of a twisted Coxeter element respectively.
Using these deformed objects, expressions for the generating function of
the multiplicities of the building block of the $S$-matrices are obtained.
The relation between the deformed version of Dorey's rule and
generalized Dorey's rule due to Chari and Pressley are discussed.
\end{quote}
\vfill
\setcounter{footnote}{0}
\end{titlepage}


\section{Introduction}

The affine Toda field theory \cite{AFZ,MOP,OT} is a theory of massive
scalar fields in 1+1 dimensions whose lagrangian has the form
\[
{\cal L}=\frac{1}{2}\partial^{\mu}\phi_a\partial_{\mu}\phi_a
-\frac{m^2}{g^2}\sum_j n_j\exp(g\alpha_j\cdot\phi).
\] 
The real coupling affine Toda field theory has infinite numbers of
commuting conserved charges which lead to diagonal $S$-matrices.
The $S$-matrices for the theories based on simply-laced algebras (and
$A_{2N}^{(2)}$ algebra) are constructed from the building block
\cite{BCDS2},
\beqn
\{x\}=\frac{(x-1)(x+1)}{(x-1+B)(x+1-B)}, \ \ \ \ 
(x)=\frac{\sinh\frac{1}{2}(\beta+i\pi x/h)}
         {\sinh\frac{1}{2}(\beta-i\pi x/h)}.
\eeqn
Here $h$ is the Coxeter number of the algebra,
$\beta$ is the rapidity difference of the two particles participating
in the scattering and $0\leq B \leq 2$ is
a function of the coupling constant conjectured to be
\cite{AFZ,BCDS2,BCDS,CM,BS} 
\beqn
B(g)=\frac{2g^2}{g^2+4\pi}.
\eeqn
In these cases, so called ADE cases, the $S$-matrix elements can be
written in the form:
\beqn
\label{ADEmult}
S_{ab}(\beta)=\prod_{x=1}^h \{x\}^{m_{ab}(x)},
\eeqn
where a non-negative integer $m_{ab}(x)$ denotes the multiplicity of
the building block $\{x\}$. Lie algebraic structure for ADE case is
well known and the multiplicity $m_{ab}(x)$ has a geometrical
representation based on the Coxeter element of the Weyl group \cite{D2}.

Quantum effects do not change the ratio of the conserved charges
characterized by a Lorentz spin. These charges are proportional to the
eigenvectors of the Cartan matrix $C_{ab}$ of the associated
non-affine Lie algebra.

For theories based on the non-simply-laced algebras, the $S$-matrices
are conjectured in \cite{DGZ} except for the dual pair $(F_4^{(1)},
E_6^{(2)})$ which was proposed in \cite{CDS}. In these cases, quantum
effects change the ratio of the conserved charges. The flow between
the dual pairs, 
$(G_2^{(1)}, D_4^{(3)})$, $(F_4^{(1)}, E_6^{(2)})$, $(C_N^{(1)},
D_{N+1}^{(2)})$ and $(B_N^{(1)}, A_{2N-1}^{(2)})$ are proposed. The
$S$-matrices and the conserved charges are known but their Lie
algebraic structure has not yet been revealed.

In \cite{D}, Dorey observed that the $S$-matrices for the dual pair
$(X_N^{(1)}, Y_M^{(l)})$ can also be expressed by the general block:
\[
\{x,y\}=\displaystyle{\frac{<x-1,y-1><x+1,y+1>}
{<x-1,y+1><x+1,y-1>}}.
\]
\[
<x,y>=<(2-B)x/2h+By/2h^{(l)\vee}>,\ \ \ \ 
<x>=\frac{\sh\frac{1}{2}(\beta+i\pi x)}{\sh\frac{1}{2}(\beta-i\pi x)},
\]
where $h^{(l)\vee}=lh^{\vee}$ is the $l$-th dual Coxeter number of
$X_N$ \cite{Ka}.  
The coupling dependent function $0\leq B\leq 2$ is proposed to be \cite{D} 
\beqn
B(g)=\frac{2g^2}{g^2+4\pi h/h^{\vee}}.
\eeqn
Here $h^{\vee}$ is the dual Coxeter number of $X_N$.
$S$-matrix elements are given by
\beqn
S_{ab}(\beta)=\prod_{x=1}^h \prod_{y=1}^{h^{(l)\vee}}
\{x,y\}^{m_{ab}(x,y)}.
\label{XYmult}
\eeqn
Here $m_{ab}(x,y)$ denotes the multiplicity of the building block $\{x,y\}$.
The multiplicity is defined for $1 \leq x \leq h$ and 
$1 \leq y \leq h^{(l)\vee}$. It is convenient to define $m_{ab}(x,y)$
for all $x, y \in {\bf Z}$ via following relations:
\beqn
m_{ab}(-x, -y) = - m_{ab}(x,y).
\eeqn
For $1 \leq x < 2h$, $1 \leq y \leq 2h^{(l)\vee}$ and $j,k \in {\bf Z}$,
\beqn
m_{ab}(x+2jh, y+2kh^{(l)\vee}) = \delta_{j,k} m_{ab}(x,y).
\eeqn

In section 2, the properties of the multiplicity are described 
from the side of the untwisted affine Lie algebra $X_N^{(1)}$. 
A Coxeter element $\sigma$ of the non-simply-laced horizontal subalgebra $X_N$
satisfies $\sigma^{-h}=1$. We introduce a ``$q$-deformation'' of the
inverse of a Coxeter element $\sigma_q^{(-)}$ such that 
$(\sigma_q^{(-)})^h = q^{2h^{(l)\vee}}$ in which $q$ is an indefinite
parameter. 
Using $\sigma_q^{(-)}$,  a ``geometrical'' expression is obtained 
for a generating function of the multiplicities. 

In section 3, the properties of the multiplicity are described from the
side of the twisted affine Lie algebra $Y_M^{(l)}$. A twisted
Coxeter element $\tilde{\sigma}$ of the simply-laced horizontal subalgebra
$Y_M$ satisfies $\tilde{\sigma}^{-h^{(l)\vee}} = 1$. We introduce a
``$p$-deformation'' of the inverse of a twisted Coxeter element
$\tilde{\sigma}_p^{(-)}$ such that
$(\tilde{\sigma}_p^{(-)})^{h^{(l)\vee}} = p^{2h}$ in which $p$ is
again a parameter. Using
$\tilde{\sigma}_p^{(-)}$, a ``geometrical'' expression is given for a
generating function of some integers which are related to the multiplicities.

In section 4, the conditions for the non-vanishing of the $n$-point
couplings are discussed. The conditions can be formulated in terms of the
$q$-deformed Coxeter element {\it or} in terms of the $p$-deformed
twisted Coxeter element. These 
conditions are equivalent to Dorey's rule \cite{D2,B} generalized to
the non-simply-laced cases by Chari and Pressley \cite{CP}. They
expressed the rule using an ordinary Coxeter element {\it and} an ordinary 
twisted Coxeter element.

In section 5, an integral representation of the $S$-matrices are given. 
Section 6 is devoted to conclusion.

\resection{$q$-deformed Coxeter element in $X_N^{(1)}$}

In this section, the multiplicity $m_{ab}(x,y)$ for the dual pair
$(X_N^{(1)}, Y_M^{(l)})$  is described using the data of
the untwisted Lie algebra $X_N^{(1)}$. The
following argument also works for simply-laced cases ($X = A, D, E$)
in which case, $Y_M^{(l)} = X_N^{(1)}$ and 
$m_{ab}(x,y) = \delta_{x,y} m_{ab}(x)$.

Let $\{\alpha_a\}$ ($a\in J=\{1, 2, \ldots, N\}$) be a set of simple
roots of the horizontal subalgebra $X_N$. The fundamental
weights $\lambda_a$ satisfy $(\lambda_a, \alpha_b)=\delta_{ab}
(\alpha_b, \alpha_b)/2$. 
Let $C_{ab}$ and $I_{ab}$ $(a, b\in J)$ denote the Cartan matrix and the 
incidence matrix respectively. 
Fix coprime positive integers
$t_a$ such that $I_{ab}t_b = I_{ba}t_a$. 
The normalization of simple roots are such
that the long roots have length squared equal to two. With this
normalization, $t_a=l(\alpha_a, \alpha_a)/2$.

For fixed $x$, the integers $m_{ab}(x,y)$ vanish for all but finite
numbers of $y$. The following matrix-valued generating function is
well defined:
\beqn
(m^q(x))_{ab} = m_{ab}^q (x) = \sum_{y\in {\bf Z}} m_{ab}(x,y)q^y.
\eeqn
Let us introduce the following matrices 
\beqn
(D_q)_{ab} = q^{t_a} \delta_{ab}, \ \ \ 
(T_q)_{ab} = [t_a]_q \delta_{ab}, \ \ \ 
(I_q)_{ab} = [I_{ab}]_q.
\eeqn
Here $[n]_q=(q^n-q^{-n})/(q-q^{-1})$ is a $q$-number.
A case-by-case analysis can be summarized as follows:
\beqn
m^q(0) = 0, \ \ \ m^q(1) = D_q T_q,
\label{initq}
\eeqn
and $m^q(x)$ satisfy recursion relations
\beqn
D_q^{-1} m^q(x+1) + D_q m^q(x-1) = I_q m^q(x).
\label{recq}
\eeqn
The initial conditions (\ref{initq}) and the recursion relation
(\ref{recq}) uniquely determine $m^q(x)$ 
for all $x \in {\bf Z}$. The solution is given by
\beqn
m^q(x) = U_{x-1}(D_q I_q; D_q^2 ) D_q T_qŽ¥,
\eeqn
for $x > 0$ and $m^q(-x)=-m^{q^{-1}}(x)$.
Here $U_x(F;G)$ is generated from the generating function
\beqn
(1 - q F + q^2 G)^{-1} = \sum_{x=0}^{\infty} U_x(F;G) q^x
\label{qChev}
\eeqn
where $F$ and $G$ are not necessary to be mutually commuting objects.
If $G=1$, $U_x(F;1)$ is the ordinary Chebyshev polynomial of the second kind.

Thus we have obtained a closed form expression of $m_{ab}(x,y)$ based
on the data of the incidence matrix only.

In order to give a ``geometrical'' expression for $m^q(x)$, 
let us consider the following $q$-deformation of the simple Weyl reflection 
corresponding to the simple root $\alpha_a$ :
\beqn
\sigma_a^q (\alpha_b) = \alpha_b - (2\delta_{ba}-[I_{ba}]_q)\alpha_a.
\eeqn
It holds that $(\sigma_a^q)^2=1$. This is almost same as the ordinary
simple reflection $\sigma_a$ except for the unique combination of the
long root $\alpha_b$ and the short root $\alpha_a$ which are adjacent
(i.e. except for $I_{ba}=l$). Only one simple reflection is deformed.

Corresponding to a Coxeter element
$\sigma = \sigma_{a_1}\sigma_{a_2}\ldots \sigma_{a_N}$, ($a_i \in J$;
all different), define a root $\phi_{a_i}$ ($a_i\in J$) by
\beqn
\phi_{a_i} = \sigma_{a_N}\sigma_{a_{N-1}}\ldots 
\sigma_{a_{i+1}} (\alpha_{a_i}),
\eeqn
and its $q$-analogue by
\beqn
\phi_{a_i}^q = \sigma_{a_N}^q\sigma_{a_{N-1}}^q\ldots 
\sigma_{a_{i+1}}^q (\alpha_{a_i}).
\eeqn
Let denote $R_a$ $(a\in J)$ the $\sigma$-orbit which contains the root 
$\phi_a$. The action of the Coxeter element $\sigma$ has a period $h$.
So the number of the element of each orbit $R_a$ equals to $h$.
Any root of $X_N$ is contained in one of $N$ orbits $R_a$ : 
$Nh = {\rm dim}X_N - N$. $\phi_a$ is the unique positive root in $R_a$
which is mapped into a negative root under the action of $\sigma$.

Let $c(a)=\pm 1$ ($a\in J$) be a two-coloring of the nodes of the
Dynkin diagram such that $I_{ab}=0$ if $a$ and $b$ have a same color,
i.e. $c(a)c(b)=1$.
It is convenient to choose the Steinberg ordering of a Coxeter element 
$\sigma$ \cite{S}:
\beqn
\sigma=\prod_{c(a)=+1}\sigma_a \prod_{c(a)=-1}\sigma_a = \sigma_+
\sigma_-.
\eeqn

Let us define 
\beqn
\sigma_{\pm}^q = \prod_{c(a)=\pm 1} \sigma_a^q.
\eeqn
It is easy to see that
\beqn
\sigma_{c(a)}^q \alpha_a = -\alpha_a, \ \ \ 
\sigma_{-c(a)}^q \alpha_a = \alpha_a + \sum_{b} [I_{ab}]_q \alpha_b.
\eeqn
Sum of these two equations leads to a $q$-analogue of Steinberg
identity \cite{S}:
\beqn
(\sigma_+^q + \sigma_-^q ) \alpha_a = \sum_b [I_{ab}]_q \alpha_b.
\eeqn
In the Steinberg ordering, $\phi_a^q$ can be expressed by
\beqn
\phi_a^q = (\sigma_{-}^q)^{(1+c(a))/2}\alpha_a.
\eeqn
There are two choice of two-coloring. 
Only one simple reflection corresponding to the short root
$\alpha_b$ which is adjacent to the long root is $q$-deformed.
If we choose the color of the node $c(b)=1$, then $\sigma_-^q =
\sigma_-$.
For this particular coloring, $\phi_a^q$ does not depend on $q$ and
coincides with $\phi_a$. But the following argument works for either
colorings.

Let us introduce a linear operator $\tau_q$ by
\beqn
\tau_q(\alpha_a) = q^{t_a} \alpha_a.
\eeqn
Then $q$-deformation of the Coxeter element and its inverse
$\sigma_q^{(\pm)}$ are defined by 
\beqn
\sigma_q^{(\pm)} = \sigma_{\pm}^q \tau_q \sigma_{\mp}^q \tau_q.
\eeqn
They have ``quasi-periodicity'':
\beqn
(\sigma_q^{(\pm)})^h = q^{2h^{(l)\vee}}.
\eeqn
For $h$ even, $(\sigma_q^{(\pm)})^{h/2} = -q^{h^{(l)\vee}}{\cal C}$
where ${\cal C}$ is the charge conjugation : 
${\cal C}(\alpha_a)=\alpha_{\bar{a}}$.
This suggests the appearance of the information of the dual algebra 
$Y_M^{(l)}$ through the $q$-deformation.
For $q=1$, $\sigma_1^{(\pm)} = \sigma^{\pm 1}$. 

The function $U_{2r}(F;G)$ ($U_{2r+1}(F;G)$) is an even (odd) function 
of $F$. $D_q$ is a diagonal matrix and 
$(I_q^{2r})_{ab}=0$ if $a$ and $b$ have different colors,
i.e. $c(a)c(b)=-1$, and $(I_q^{2r+1})_{ab}=0$ 
if $a$ and $b$ have the same color, i.e. $c(a)c(b)=1$.
Hence $m_{ab}^q(2r+\epsilon_{ab})=0$ where $\epsilon_{ab}=(c(a)-c(b))/2$.
It is easy to see that $2(\lambda_b, \phi_a^q)/(\alpha_b, \alpha_b)$
equals to $0$, $\delta_{ab}$  
and $[ I_{ab} ]_q$ for $\epsilon_{ab} = -1, 0, 1,$ respectively.
Making use of this relation and 
\beqn
( \sigma_q^{(-)} + q^{2t_a} ) \phi_a^q =
\sum_b q^{ ( t_a + t_b )( 1 - c(a) )/2 } [ I_{ab} ]_q 
( \sigma_q^{(-)} )^{ ( 1 + c(a) )/2 } \phi_b^q,
\eeqn
it is possible to show that the multiplicity
$m_{ab}(2r+1+\epsilon_{ab})$ has an expression
\beqn
m_{ab}^q(2r+1+\epsilon_{ab})=q^{\kappa_{ab} + t_b} 
\frac{2}{(\alpha_b, \alpha_b)} [t_b]_q
\ (\lambda_b , (\sigma_q^{(-)})^r\phi_a^q),
\label{qmab}
\eeqn
which indeed satisfies the recursion relation (\ref{recq}) with the
correct initial conditions (\ref{initq}).
Here $\kappa_{ab} = ( 1 + c(a) ) t_a/2 - ( 1 + c(b) ) t_b/2$.
Thus we have obtained a ``geometrical'' expression using the
$q$-deformed Coxeter element. This is one of the main results of this paper.

For the simply-laced cases, $\kappa_{ab} = \epsilon_{ab}$ and
$\sigma_q^{(\pm)}=q^2\sigma^{\pm 1}$. The above relation becomes
\[
m_{ab}^q(2r+1+\epsilon_{ab})=q^{2r+1+\epsilon_{ab}}
(\lambda_b , \sigma^{-r}\phi_a), \ \ \ r \in {\bf Z}.
\]
This implies that $m_{ab}(x,y)=\delta_{x,y}m_{ab}(x)$ and 
\beqn
m_{ab}(2r+1+\epsilon_{ab})=(\lambda_b , \sigma^{-r}\phi_a), 
\ \ \ r \in {\bf Z},
\label{DG}
\eeqn
which is Dorey's geometrical formula \cite{D2,D3}.

\resection{$p$-deformed Twisted Coxeter element in $Y_M^{(l)}$}

The conjectured strong-weak duality for the pair 
$(X_N^{(1)}, Y_M^{(l)})$ implies that the strong coupling
affine Toda field theory based on $X_N^{(1)}$ can be described by the 
weak coupling affine Toda field theory based on $Y_M^{(l)}$.
It should be possible to characterize the multiplicity of the
$S$-matrix using the data of twisted algebra $Y_M^{(l)}$. 

Dorey observed that the twisted Coxeter element $\tilde{\sigma}$ plays
crucial role in the non-simply-laced cases \cite{D3}.
In this section, another ``geometrical'' expression for the
multiplicity is given using an analogue of the twisted Coxeter
element of the simply-laced horizontal subalgebra $Y_M$.

The twisted algebra $Y_M^{(l)}$ can be constructed from the 
simply-laced horizontal subalgebra $Y_M$ and the Dynkin diagram automorphism 
$\dot{\omega}$ such that $\dot{\omega}^l=1$. 

The incidence matrix of $Y_M$ is denoted by $\tilde{I}_{ij}$ where 
$i, j \in \tilde{J}=\{1,2,\ldots, M\}$. 
Let $\{ \tilde{\alpha_i} \}_{ i \in \tilde{J} }$ and 
$\{ \tilde{\lambda}_i \}_{ i \in \tilde{J} }$ be a set of simple roots and a
set of fundamental weights of $Y_M$ such that 
$(\tilde{\lambda}_i, \tilde{\alpha}_j) = \delta_{ij}$.
(The simple roots are normalized to have length square equal to two.) 
We denote by $l_i$ the smallest
positive integer such that $\dot{\omega}^{l_i}(i)=i$.
Each $\dot{\omega}$-orbit
\[
\{ i, \dot{\omega}(i), \ldots, \dot{\omega}^{l_i-1}(i) \}
\]
corresponds to a node of the Dynkin diagram of the dual Lie algebra
$X_N^{\vee}$ of $X_N$. The (non-affine) Lie algebra $X_N$ is
self-dual except for $B_N$ and $C_N$ which are dual each other.
The identification of $\dot{\omega}$-orbits and the nodes of the Dynkin
diagram of $X_N^{\vee}$ is nothing but the folding of the Dynkin
diagram of $Y_M$ by the automorphism $\dot{\omega}$.

We take the representatives of the $\dot{\omega}$-orbits as
$J=\{1,2,\ldots, N\}$ and identify with the particle species. Let
${\bf r}(i)=\pm 1$ according to $i$ is 
a representative or not. For the nodes of the Dynkin diagram of $Y_M$, 
we give a two-coloring $c(i) = \pm 1$ such that $\tilde{I}_{ij} = 0$
if $c(i)c(j)=1$. For $l>1$, there exists a unique node $i$ with $ l_i = 1 $, 
which is connected to a node $j$ with $l_j = l$. Let $c_0 = c(i)$.
For later convenience, let $c_0$ be a finite constant for trivial
automorphism ($l=1$). 

The automorphism $\omega$ among simple roots $\{\tilde{\alpha}_i\}_{i\in
\tilde{J}}$ of $Y_M$ is defined through the Dynkin diagram
 automorphism $\dot{\omega}$ as follows:
\beqn
\omega(\tilde{\alpha}_i)
=\tilde{\alpha}_{\dot{\omega}(i)}, \ \ \ i\in \tilde{J}.
\eeqn

As was discussed in the previous section, except for the self-dual
pair $(A_{2N}^{(2)}, A_{2N}^{(2)})$, the multiplicity matrix
$m_{ab}^q(x)$ (\ref{qmab}) contains a factor $[t_b]_q$. This 
suggests that the multiplicity can be explained by folding. The
$A_{2N}^{(2)}$ case are excluded in the argument below and will be
discussed at the end of this section.

Under the action of the twisted Coxeter element $\tilde{\sigma}$, 
roots of $Y_M$ are grouped into $\tilde{\sigma}$-orbits. 
The twisted Coxeter element has a period $h^{(l)\vee}$. 
For $l>1$, this is longer than the period of a Coxeter element of $Y_M$. 
Every $\tilde{\sigma}$-orbit contains $h^{(l)\vee}$ elements. 
So the number of the $\tilde{\sigma}$-orbits equals to $N$ which is
smaller than $M$, the number of orbits of the Coxeter element.

If we use a particular element of the $\tilde{\sigma}$-orbits and 
the fundamental weight $\tilde{\lambda}_i$, 
we will have an $N$ by $M$ matrix. But the multiplicity $m(x,y)$ is an
$N$ by $N$ matrix. So we first introduce an auxiliary $N$ by $M$
rectangular matrix $n(x,y)$ whose elements take values in integers, 
then relate $n(x,y)$ to $m(x,y)$.

Let us introduce a rectangular matrix $n_{ai}(x,y)$ 
$(a\in J, i\in \tilde{J}$ and $x,y \in {\bf Z})$ using 
the following generating matrix of $n_{ai}(x,y)$:
\beqn
(n^p(y))_{ai} = n_{ai}^p(y) = \sum_{x \in {\bf Z} } n_{ai}(x,y) p^x.
\eeqn
The matrix $n^p(y)$ is defined by a recursion relation:
\beqn
n^p(y + 1) W = n^p(y) F_p - n^p(y - 1) G_p,
\eeqn
with initial conditions $n^p(0) = 0$ and 
$n_{ai}^p(1) = p \delta_{a,\dot{w}(i)}$.
Here $ F_p = p \tilde{I}_o {\bf P}_r $ and 
$ G_p = \tilde{D}_p - p W ( \tilde{I} - \tilde{I}_o) {\bf P}_r $.
The explicit form of these matrices are given by
\beqa
( \tilde{I}_o )_{ij} &=& 
    ( 1 - ( 1 - \delta_{l_i,1} )( 1 - \delta_{l_j,1} ) ) \tilde{I}_{ij}, \ \ \ 
( {\bf P}_r )_{ij} = \delta_{ij} ( 1 + {\bf r}(j) )/2, \nn 
( W )_{ij} &=& \delta_{\dot{w}(i),j}, \ \ \ 
( \tilde{D}_p )_{ij} = {\bf r}(i) p^{ 1 + {\bf r}(i) }\delta_{ij}.
\eeqa
Using $U_x(F;G)$ (\ref{qChev}), $n^p(y)$ can be expressed in the
following form:
\beqn
n^p(y) = n^p(1) U_{y-1}( F_p W^{-1}; G_p W^{-1} ).
\eeqn

The multiplicity matrix $m_{ab}(x,y)$ is related to 
$n_{ai}(x,y)$ in the following form:
\beqn
m_{ab}(x,y)=\sum_{k=0}^{l_b-1}n_{a,\dot{w}^k(b)}(x,y),\ \ \ \ 
a,b\in J.
\label{mton}
\eeqn
Thus we have obtained another closed form expression of $m_{ab}(x,y)$
based on the data of the incidence matrix and the automorphism.

In order to give a ``geometrical'' expression,
let us consider a deformation of the twisted Coxeter element
$\tilde{\sigma}$ of $Y_M$ \cite{S,Sp}: 
\beqn
\tilde{\sigma}=\tilde{\sigma}_{a_1}\tilde{\sigma}_{a_2}\ldots
\tilde{\sigma}_{a_N}\omega, \ \ \ \ a_k\in J.
\eeqn
A root $\tilde{\phi}_{a_k}$ is defined by
\beqn
\tilde{\phi}_{a_k}
=\omega^{-1} \tilde{\sigma}_{a_N} \tilde{\sigma}_{a_{N-1}} \ldots 
\tilde{\sigma}_{a_{k+1}} (\tilde{\alpha}_{a_k}).
\eeqn
It is convenient to take the Steinberg ordering:
\beqn
\tilde{\sigma} = \tilde{\sigma}_+ \tilde{\sigma}_- \omega, \ \ \ 
\tilde{\sigma}_{\pm} = \prod_{c(a)=\pm 1,\ a \in J} \tilde{\sigma}_a.
\eeqn
Let us define a linear operator $\tilde{\tau}_p$ acting on simple
roots of $Y_M$ by
\beqn
\tilde{\tau}_p (\tilde{\alpha}_i) = p^{1 + {\bf r}(i)} \tilde{\alpha}_i.
\eeqn
Then a ``$p$-deformation'' of the inverse of the twisted Coxeter element is
defined by 
\beqn
\tilde{\sigma}_p^{(-)} 
= w^{-1} \tilde{\sigma}_- \tilde{\tau}_p \tilde{\sigma}_+,
\eeqn
which coincides with the inverse of the ordinary twisted Coxeter
element at $p=1$. 
It has ``quasi-periodicity'':
\beqn
(\tilde{\sigma}_p^{(-)})^{h^{(l)\vee}} = p^{2h}.
\eeqn
For $h^{(l)\vee}$ even, 
$(\tilde{\sigma}_p^{(-)})^{h^{(l)\vee}/2} = -p^{h}{\cal C}$ where 
${\cal C}$ is the charge conjugation : 
${\cal C}(\tilde{\alpha}_i)=\tilde{\alpha}_{\bar{i}}$.
The information of the dual algebra emerges through a $p$-deformation.
We find that $n^p(y)$ can be expressed in the following form:
\beqa
n_{ai}^p(2r+\tilde{\kappa}_{ai})&=&0, \nn
n_{ai}^p(2r+1+\tilde{\kappa}_{ai})&=&
p^{1+\epsilon_{ai}}(\tilde{\lambda}_i,
(\tilde{\sigma}_p^{(-)})^r\tilde{\phi}_a), \ \ \ \ 
r\in {\bf Z}.
\label{ngeo}
\eeqa
Here $\tilde{\kappa}_{ai}=(c_0 + c(a))l_a/2 - (c_0 + c(i))l_i/2$ and 
$\epsilon_{ai}=(c(a)-c(i))/2$. For $Y_M^{(l)} = D_4^{(3)},
D_{N+1}^{(2)}$, we can see that $\tilde{\kappa}_{ai}$ equals to 
$\epsilon_{ai}$.
Combining this expression and eq.(\ref{mton}), we have another
``geometrical'' expression using $p$-deformed twisted Coxeter
element. This is also one of the main results of the paper.

The expression 
(\ref{ngeo}) also give the correct answer for 
trivial automorphism, i.e. $l=1$, $J=\tilde{J}$. 
Because $l_a=1$ for all $a$, the value of $c_0$ is irrelevant for
$\tilde{\kappa}_{ab}$.
$\tilde{\kappa}_{ab}$ coincides with $\epsilon_{ab}$ and 
the inverse of ($p$-deformed) twisted Coxeter element becomes
($p^2$ times) inverse of ordinary Coxeter element. 
Thus $m_{ab}(x,y)=n_{ab}(x,y)$ and 
\[
n^p_{ab}(2r+1+\epsilon_{ab})=
p^{2r+1+\epsilon_{ab}}(\tilde{\lambda}_b, \sigma^{-r}\tilde{\phi}_a),
\ \ \ \ r\in {\bf Z},
\]
which leads to the Dorey's geometrical formula (\ref{DG}).

In the rest of this section, let us discuss the self-dual case
$A_{2N}^{(2)}$. 
In this case, replace the second dual Coxeter number $h^{(2)\vee}$ in
eq. (\ref{XYmult}) by the dual Coxeter number $h^{\vee}=2N+1$ and
$m_{ab}(x,y)=\delta_{x,y}m_{ab}(x)$. Here the multiplicity $m_{ab}(x)$
is given by eq.(\ref{ADEmult}) with $h=2N+1$. We have $t_a=1$. The multiplicity
$m_{ab}^q(x)$ satisfies the recursion relation (\ref{recq}) if we replace
$I_{ab}$  by the ``generalized incidence matrix'' \cite{KM}
which is the matrix obtained from the incidence matrix of $A_N$ algebra
$I^{(A_N)}$ by replacing the zero in 
the last entry along the diagonal by one.  The multiplicity can also
be expressed using the twisted Coxeter element by
\beqn
m_{ab}( r + \epsilon_{ab} ) 
= ( \tilde{\lambda}_b, \tilde{\sigma}^{-r} \tilde{\phi}_a)
= ( \tilde{\lambda}_{\dot{w}(b)}, \tilde{\sigma}^{-(r-1)}\tilde{\phi}_a).
\eeqn

\resection{Dorey's Rule}

In section 2 and 3, we introduced the deformation of the Coxeter
element and twisted Coxeter element. In this section, we discuss the
condition when the fusion process occurs in terms of these deformed
objects. We also discuss the non-vanishing condition of (quantum)
$n$-point couplings.

In terms of the multiplicity $m_{ab}(x,y)$, the fusing $a \times b
\rightarrow \bar{c}$ occurs if it holds that
\beqn
m_{ad}( x + \bar{u}_{ac}^b(0), y + \bar{u}_{ac}^b(2) ) + 
m_{bd}( x - \bar{u}_{bc}^a(0), y - \bar{u}_{bc}^a(2) ) =
m_{\bar{c}d}( x , y ),
\eeqn
for some integers $\bar{u}(0)$ and $\bar{u}(2)$.
We can treat above relation either from the viewpoint of $X_N^{(1)}$
or from the viewpoint of $Y_M^{(l)}$.

We first discuss the fusion process from the side of $X_N^{(1)}$.
Making use of $m^q(x)$, the above relation can be expressed as follows:
\beqn
q^{-\bar{u}_{ac}^b(2)} m_{ad}^q( x + \bar{u}_{ac}^b(0)) +
q^{\bar{u}_{bc}^a(2)} m_{bd}^q( x - \bar{u}_{bc}^a(0) ) = 
m_{\bar{c}d}^q(x).
\eeqn
This holds for any particle species $d$, so we can see that
the fusing $a \times b \rightarrow \bar{c}$ occurs if and only if
there exist some integers $r_1, r_2, u_1$ and $u_2$ such that
\beqn
q^{u_1}(\sigma_q^{(-)})^{r_1} \phi_a^q + 
q^{u_2}(\sigma_q^{(-)})^{r_2} \phi_b^q = \phi_{\bar{c}}^q.
\eeqn

Next, we discuss the fusion process from the side of $Y_M^{(l)}$.
Another way of expressing the fusing is
\beqn
\sum_{k=0}^{l_d - 1}\left\{
p^{-\bar{u}_{ac}^b(0)} n_{a,\dot{\omega}^k(d)}^p( y + \bar{u}_{ac}^b(2)) +
p^{\bar{u}_{bc}^a(0)} n_{b,\dot{\omega}^k(d)}^p( y - \bar{u}_{bc}^a(2) ) - 
n_{\bar{c},\dot{\omega}^k(d)}^p(y)\right\} = 0.
\eeqn
This must hold for all $d$. Hence 
the fusing occurs if and only if there exist some integers $r_1',
r_2', u_1'$ and $u_2'$ such that
\beqn
p^{u_1'}(\tilde{\sigma}_p^{(-)})^{r_1'} \tilde{\phi}_a +
p^{u_2'}(\tilde{\sigma}_p^{(-)})^{r_2'} \tilde{\phi}_b =
\tilde{\phi}_{\bar{c}}.
\eeqn
These relations can be easily generalized to the conditions for
$n$-point couplings among the particles $a_1, a_2, \ldots, a_n$. The
$n$-point coupling is non-vanishing if and 
only if there exist integers $r_i$, $u_i$ ($i=1, \ldots,n$) such that
\beqn
q^{u_1}(\sigma_q^{(-)})^{r_1} \phi_{a_1}^q +
q^{u_2}(\sigma_q^{(-)})^{r_2} \phi_{a_2}^q +
\cdots +
q^{u_n}(\sigma_q^{(-)})^{r_n} \phi_{a_n}^q = 0.
\label{Dnq}
\eeqn
The deformation parameter $q$ carries the information of $Y_M^{(l)}$,
the dual of $X_N^{(1)}$. If one sets $q=1$, the $q$-deformed Coxeter
element reduces to the ordinary Coxeter element and the information of 
the dual part is lost.

Alternatively, in terms of $\tilde{\sigma}_p^{(-)}$, it is possible to 
say that the $n$-point couplings does not vanish if and only if there
exist some integers $r_i'$, $u_i'$ ($i=1, \ldots,n$) such that
\beqn
p^{u_1'}(\tilde{\sigma}_p^{(-)})^{r_1'} \tilde{\phi}_{a_1} +
p^{u_2'}(\tilde{\sigma}_p^{(-)})^{r_2'} \tilde{\phi}_{a_2} +
\cdots +
p^{u_n'}(\tilde{\sigma}_p^{(-)})^{r_n'} \tilde{\phi}_{a_n} =0.
\label{TDnp}
\eeqn
The deformation parameter $p$ carries the information of $X_N^{(1)}$,
the dual of $Y_M^{(l)}$. If one sets $p=1$, the $p$-deformed twisted Coxeter
element reduces to the ordinary twisted Coxeter element and the
information of the dual part is lost. 

Generalized Dorey's rule \cite{CP} is that
the $n$-point coupling does not vanish if
both eq.(\ref{Dnq}) with $q=1$ and eq.(\ref{TDnp}) with $p=1$ holds.
Both of the information comes from $X_M^{(1)}$ and $Y_M^{(l)}$ are
necessary to judge if the $n$-point coupling vanish or not.
This is also correct for the generalized bootstrap principle
\cite{CDS} which requires the information on the sign of the residue
of the poles for all range of the coupling constant.

\resection{Integral representations of the $S$-matrix}

In this section, an integral representation of the $S$-matrix is given 
using the data of $X_M^{(1)}$. In expressing the phase shift, an
inverse of a matrix $C(k)$ appears. It has very universal form
(\ref{gC}).

For the dual pair $(X_N^{(1)}, Y_M^{(l)})$, 
the recursion relation (\ref{recq}) leads to an integral
representation of the $S$-matrix.
The multiplicity $m(x,y)$ has a property: 
\[
m(x,y)
={\cal C}m(h-x,h^{(l)\vee}-y)= m(h-x,h^{(l)\vee}-y){\cal C},
\] 
where ${\cal C}_{ab}=\delta_{a\bar{b}}$ is the charge conjugation matrix.
Then $m^q(h-1)={\cal C}q^{h^{(l)\vee}}m^{q^{-1}}(1)
=q^{h^{(l)\vee}}{\cal C}D_q^{-1}T_q$. From $m^q(h)=0$ and
eq.(\ref{recq}), we have $m^q(h+1)=-q^{h^{(l)\vee}}{\cal C}D_q T_q$.
Thus we have
\beqn
\sum_{x=1}^{h}m^q(x)p^x=(pD_q+p^{-1}D_q^{-1}-I_q)^{-1}
(1_N+p^h q^{h^{(l)\vee}}{\cal C})T_q.
\eeqn
Here $1_N$ is the $N$ by $N$ identity matrix.
Using the above relation, we obtain an integral representation of the
$S$-matrix:
\beqn
S_{ab}(\beta)=(-1)^{\delta_{ab}}
\exp\left\{-2\int_0^{\infty}\frac{dk}{k}\sin k\beta
\left(K(k)(\breve{C}(k)^{-1})_{ab}-\delta_{ab}\right)\right\},
\eeqn
where
\beqn
K(k)=4\sh\frac{(2-B)}{2h}\pi k \ \sh\frac{B}{2h^{(l)\vee}}\pi k,
\eeqn
\beqn
\label{tC}
\breve{C}_{ab}(k)=[t_a]_{q(k)}^{-1}C_{ab}(k), \ \ \ 
q(k)=\exp(B\pi k/2h^{(l)\vee}),
\eeqn
\beqn
\label{gC}
C_{ab}(k)=\left(\delta_{ab}2\cosh\left(\frac{(2-B)}{2h}+
\frac{B}{2h^{(l)\vee}}t_a\right)\pi k - [I_{ab}]_{q(k)}\right).
\eeqn
Using
\beqn
[I_{ab}]_q [t_b]_q = [I_{ba}]_q [t_a]_q,
\eeqn
we can see that $\breve{C}_{ab}(k)$ is a symmetric matrix.
For $k=0$ or $B=0$, the matrix $C_{ab}(k)$ coincides with $C_{ab}$,
the Cartan matrix of $X_N$. 

For ADE cases ($l=1$), $C_{ab}(k)=\delta_{ab}2\cosh(\pi k/h) -
I_{ab}$, which is the same as the case of the minimal ADE $S$-matrix
\cite{Z}\footnote{
For minimal ADE $S$-matrix, the kernel function $K(k)$ is given by 
$2\cosh(\pi k/h)$. Then it is known that
$K(k)C(k)^{-1}-1_N=I(2\cosh(\pi k/h)1_N-I)^{-1}$ \cite{Z}.
For $k=0$, this relation was observed by Klassen and Melzer \cite{KM}.}.
Interestingly, for $l \neq 1$, $C(k)$ depends on $B$ but ${\rm det}
C(k)$ does not.
Explicitly, ${\rm det} C(k)$ equals to 
$\cosh (\pi k/2) / \cosh (\pi k/6)$
for $(G_2^{(1)}, D_4^{(3)})$ and $(F_4^{(1)}, E_6^{(2)})$,
$2\cosh (\pi k/2)$
for $(C_N^{(1)}, D_{N+1}^{(2)})$ and $(B_N^{(1)}, A_{2N-1}^{(2)})$.
For general $l$, we can evaluate it at $B=0$. Recall that the
incidence matrix of $X_N$ has the right (left) eigenvector $q_a^{(s)}$ 
($q_a^{(s)\vee}$) with the eigenvalue $2\cos(\pi s/h)$ ($s\in E$). 
Here $E$ is the set of the exponents of $X_N$. Then 
\beqa
\label{detC}
{\rm det}C(k)&=&{\rm det}\left(2\cosh(\pi k/h)1_N - I \right), \nn
&=&
\prod_{s\in E}\left(2\cosh(\pi k/h)-
2\cos(\pi s/h)\right).
\eeqa
Thus at $k=i\times {\rm (affine \ exponent)}$, it vanishes.
Let $\tilde{E}$ be the set of the affine exponents : 
$\tilde{E}=\{s+rh | s\in E, r\in {\bf Z}_{\geq 0}\}$.
The matrix $C(is)$ ($s\in \tilde{E}$) has a right (left) eigenvector
$Q_a^{(s)}$ ($Q_a^{(s)\vee}$) with zero eigenvalue. 
\beqn
\sum_b C_{ab}(is)Q_b^{(s)}=0, \ \ \ \ \sum_a Q_a^{(s)\vee}
C_{ab}(is)=0.
\eeqn
Using Kramer's formula, we can determine the ratios
$Q_a^{(s)}/Q_b^{(s)}$ and $Q_a^{(s)\vee}/Q_b^{(s)\vee}$.
If we denote the cofactor of $C_{ab}(k)$ by $A_{ab}(k)$ then it
factorizes at $k=is$:
\beqn
A_{ab}(is)=Q_a^{(s)\vee}Q_b^{(s)} d_s.
\eeqn
Here $d_s$ is independent of the particle species.

It is possible to choose
\beqn
Q_a^{(s)}=[t_a]_{q(is)} Q_a^{(s)\vee}.
\eeqn
Then the $S$-matrix can be written in the form:
\beqn
S_{ab}(\beta)=(-1)^{\delta_{ab}}
\exp\left(\epsilon_{\beta}\sum_{s\in \tilde{E}}
Q_a^{(s)}Q_b^{(s)}\lambda_s
e^{-s\epsilon_{\beta}\beta} \right),
\eeqn
where $\epsilon_{\beta}=sgn \ \beta$. From the above expression,
the contribution of modes with Lorentz spin $s$ to the phase
shift is easily read off. Only the modes whose spin equals to the
exponent of the affine algebra appears. $\lambda_s$ is given by
\[
\lambda_s=\frac{h}{s}\frac{K(is)}{\sin(\pi s/h)}
d_s\prod_{
\scriptstyle s'\in E \atop\scriptstyle s'\neq s \ {\rm mod}\ h}
\left(2\cos(\pi s/h)-2\cos(\pi s'/h)\right)^{-1}.
\]
Especially for the ADE cases, $Q_a^{(s)}Q_b^{(s)}\lambda_s$ has a simple form:
\[
Q_a^{(s)}Q_b^{(s)}\lambda_s=
q_a^{(s)}q_b^{(s)}\frac{h}{s}\frac{K(is)}{\sin(\pi s/h)}.
\]
Here the eigenvectors of the incidence matrix are normalized
such that $\sum_{s\in E}q_a^{(s)}q_b^{(s)}=\delta_{ab}$.

From the consistency with the bootstrap, the right null vector
$Q_a^{(s)}$ is nothing but 
the eigenvalue of the quantum conserved charges.
At $B=0$, $Q_a^{(s)}$ is proportional to the right eigenvector of the Cartan
matrix of $X_N$. This is consistent with \cite{F,FLO}.

Remark: The matrix $C_{ab}(k)$ (\ref{gC}) is very universal.
For example, $C_{ab}(k)$ for $B=2$ appears in an integral
representation of the scalar factor of the $S$-matrix of the
$X_N\times X_N$-invariant sigma model \cite{OW}.
It is known that at $B=2$, 
the mass ratios are equal to those of $X_N\times X_N$-invariant sigma
model \cite{BCDS2}. The appearance of $C(k)$ or $\breve{C}(k)$ in the
model can be
easily seen if we use a slight different expression for $\breve{C}(k)$:
\beqn
\label{CB2}
\breve{C}_{ab}(k)
=\frac{\sh\left(2\breve{C}_{ab}\frac{B\pi k}{2h^{(l)\vee}}\right)}
{\sh\left(2\frac{B\pi k}{2h^{(l)\vee}}\right)}(1-\delta_{ab})+
\delta_{ab}
\frac{2\ch\left(\frac{(2-B)}{2h^{(l)}}+\frac{B}{2h^{(l)\vee}}t_b
\right)\pi k 
\sh\left(\frac{B\pi k}{2h^{(l)\vee}}\right)}
{\sh\left(t_b \frac{B\pi k}{2h^{(l)\vee}}\right)},
\eeqn
where $\breve{C}_{ab}=t_a^{-1}C_{ab}$.
If we identify $\pi k/h^{(l)\vee}$ with $\omega/2$ 
in \cite{OW} and take the kernel function $K(k)$ to be
$2\ch\frac{1}{2}\omega$ except for the case $(B_N^{(1)},
A_{2N-1}^{(2)})$. For this case, $\pi k/h^{(l)\vee}=\omega/4$ and
$K(k)=\ch\frac{1}{2}\omega/\ch\frac{1}{4}\omega$.
Then our $K(k)(\breve{C}(k))^{-1}$ at $B=2$ is identical with
$C(\omega)^{-1}$ in \cite{OW}.

\resection{Conclusion}

In this paper, we introduced the deformation of the (twisted) Coxeter
element. The indeterminates $q$ ($p$) carries the information of the
dual algebra. The ``geometrical'' formula for the generating functions
of the multiplicities are obtained.

A generalized Dorey's rule can be written in terms of the Coxeter
element {\it and} the twisted Coxeter element. 
We find that the equivalent condition can be written in terms of the
$q$-deformed Coxeter element {\it or} $p$-deformed Coxeter element.

Another attempt to explain the structure of the $S$-matrix 
of the non-simply laced theories and the twisted theories by folding
is found in \cite{P}.

\section*{Acknowledgments}
I would like to thank Prof. R. Sasaki for careful reading of the manuscript.
This work is partially supported by the JSPS Research Fellowships for
Young Scientists.


\appendix
\resection{The $S$-matrix in the general building block}

In this appendix, we give the $S$-matrix elements for the
non-self-dual pair $(X_N^{(1)}, Y_M^{(l)})$ in the general building
block.
Following \cite{D}, we introduce the following objects:
\[
\{x,y\}_2=\{x,y-1\}\{x,y+1\},
\]
\[
\{x,y\}_3=\{x,y-2\}\{x,y\}\{x,y+2\},
\]
\beqn
\label{defaxyb}
_2\{x,y\}_2=\{x-1,y\}_2\{x+1,y\}_2.
\eeqn
For the affine Toda theories, cancellations of zeros and poles occur
and the building block is given by
\[
_a\{x,y\}_b=\displaystyle{\frac{<x-a,y-b><x+a,y+b>}{<x-a,y+b><x+a,y-b>}}.
\]
The crossing property of the building block is
\[
_a\{x,y\}_b(i\pi-\beta)=_a\{h-x,h^{(l)\vee}-y\}_b(\beta).
\]
It is convenient to set
\[
_a[x,y]_b=_a\{x,y\}_b\times{\rm crossing}=
_a\{x,y\}_b\times_a\{h-x,h^{(l)\vee}-y\}_b.
\]
The suffix will be omitted if it equals to one.

\subsection{The case $(G_2^{(1)}, D_4^{(3)})$}

$h=6$ and $h^{(3)\vee}=12$.
\beqa
S_{11}&=&[1,1]\{3,6\}_2, \nonumber \\
S_{12}&=&[2,4]_3, \nonumber \\
S_{22}&=&[1,3]_3[3,5]_3. \nonumber
\eeqa

\subsection{The case $(F_4^{(1)}, E_6^{(2)})$}

$h=12$ and $h^{(2)\vee}=18$.
\beqa
S_{11}&=&[1,1][5,7], \nonumber \\
S_{12}&=&[4,6]_2, \nonumber \\
S_{13}&=&[2,2][4,6]\{6,9\}_2, \nonumber \\
S_{14}&=&[3,4]_2[5,8]_2, \nonumber \\
S_{22}&=&[1,2]_2[5,8]_2, \nonumber \\
S_{23}&=&[3,5]_2[5,7]_2, \nonumber \\
S_{24}&=&[2,4]_2[4,6]_2[6,8]_2, \nonumber \\
S_{33}&=&[1,1][3,4]_2[5,8]_2[5,7], \nonumber \\
S_{34}&=&[2,3]_2[4,5]_2[4,7]_2[6,9]_2, \nonumber \\
S_{44}&=&[1,2]_2[3,4]_2\left(_2[4,6]_2\right)
      \left([5,8]_2\right)^2. \nonumber
\eeqa

\subsection{The case $(C_N^{(1)}, D_{N+1}^{(2)})$}

$h=2N$ and $h^{(2)\vee}=2N+2$.
\[
S_{ab}=\prod_{
\scriptstyle p=|a-b|+1 \atop\scriptstyle {\rm step}\ 2}^{a+b-1}
[p,p].
\]
For $a+b>N$, the cancellation of poles and zeros occur and
the reduced expression is
\[
S_{ab}=\prod_{
\scriptstyle p=|a-b|+1 \atop\scriptstyle {\rm step}\ 2}^{2N-a-b-1}
[p,p]
\prod_{
\scriptstyle p=N+1-a-b \atop\scriptstyle {\rm step}\ 2}^{a+b-N-1}
\{N-p,N+1-p\}_2.
\]

\subsection{The case $(B_{N}^{(1)}, A_{2N-1}^{(2)})$}

$h=2N$ and $h^{(2)\vee}=4N-2$.
\beqa
S_{ab}&=&\prod_{
\scriptstyle p=|a-b|+1 \atop\scriptstyle {\rm step}\ 2}^{a+b-1}
[p, 2p]_2, \nonumber \\
S_{aN}&=&\prod_{
\scriptstyle p=1 \atop\scriptstyle {\rm step}\ 2}^{2a-1}
\{N-a+p,2N-1-2a+2p\}_2, \nonumber \\
S_{NN}&=&\prod_{
\scriptstyle p=1-N \atop\scriptstyle {\rm step}\ 2}^{N-1}
\{N-p, 2N-1-2p\}. \nonumber
\eeqa

\end{document}